\documentclass[prb,twocolumn,nobibnotes,groupedaddress]{revtex4}
 
\usepackage[normalem]{ulem}
\usepackage[hidelinks]{hyperref}
\usepackage[version=3]{mhchem} % Formula subscripts using \ce{}
\usepackage{braket}
\usepackage{xcolor}
\usepackage{graphicx}
\DeclareMathOperator*{\argminA}{arg\,min}
\begin{document}

\title{Analysis of fourth-, fifth-, and infinite-order triple excitations in unitary coupled cluster theory}

\thanks{This manuscript has been authored by UT-Battelle, LLC, under Contract No.~DE-AC0500OR22725 with the U.S.~Department of Energy. The United States Government retains and the publisher, by accepting the article for publication, acknowledges that the United States Government retains a non-exclusive, paid-up, irrevocable, world-wide license to publish or reproduce the published form of this manuscript, or allow others to do so, for the United States Government purposes. The Department of Energy will provide public access to these results of federally sponsored research in accordance with the DOE Public Access Plan.}

\begin{abstract}
In this work, we introduce a correction to the unitary coupled cluster method with single and double excitations (UCCSD) that incorporates the effects of missing triple excitations through a treatment that is correct through fifth-order in many-body perturbation theory (MBPT), which we refer to as UCCSD[T-5]. We then benchmark the performance of UCCSD[T-5] alongside the previously developed fourth-order UCCSD[T], comparing both against the infinite-order treatment of triples in UCCSDT as well as full configuration interaction (FCI). Two key findings emerge from this analysis. First, the fourth-order correction in UCCSD[T] consistently provides the closest agreement with FCI in estimating ground state energies, outperforming both UCCSD[T-5] and UCCSDT.  Second, the inclusion of fifth-order corrections as in UCCSD[T-5] largely recovers the infinite-order triples limit in UCCSDT. With the growing interest in UCC ansätze for quantum computing and the constraints imposed by current quantum hardware, these results underscore the potential of using classically computed perturbative corrections within UCC theory to salient triple excitation effects without requiring the additional quantum resources as would be demanded by the UCCSDT ansatz.
  
\end{abstract}

\author{Zachary W. Windom and Daniel Claudino\footnote{\href{mailto:claudinodc@ornl.gov}{claudinodc@ornl.gov}}}
\affiliation{Quantum Information Science Section,\ Oak\ Ridge\ National\ Laboratory,\ Oak\ Ridge,\ TN,\ 37831,\ USA}
\maketitle

\section{Introduction}
The exact solution to the time-independent, non-relativistic electronic Schr\"{o}dinger equation represents a ``holy grail" in quantum chemistry. From a methodological point of view, this is achieved by full configuration interaction (FCI), which enables predictive resolution of all ground and excited state properties, exact within the limits of basis set incompleteness and the validity of the Born-Oppenheimer approximation. The computational scaling of FCI is exponential in the dimensiion of the underlying Hilbert space of all electron excitations out of the reference space, thereby limiting its routine use to small collections of light atoms. 

With the recent emergence of quantum computing as a potential driver in quantum chemistry inquiry, ans\"{a}tze based on the unitary coupled cluster (UCC) theory\cite{bartlett1989some,bartlett1989alternative,szalay1995alternative,watts1989unitary,liu2018unitary,anand2022quantum} have become increasingly popular. This is in no small part due to such ans\"{a}tze having a well-defined route to achieving FCI by modulating the maximal electronic excitation rank, embodied by the anti-hermitian cluster operator $\tau=T-T^\dagger$, where $T$ is the cluster operator as defined by conventional coupled cluster (CC) theory.\cite{bartlett2007coupled,shavitt2009many} 
This in line with the fact that consideration of higher-rank wavefunction amplitudes is absolutely necessary in situations involving ``strongly correlated" electrons and for high-accuracy electronic structure applications.\cite{paldus1982cluster,paldus1984degeneracy,piecuch1990coupled,windoma2024t2,windom2024ultimate,thorpe2024factorized} Given the backdrop against which UCC is being posited, that is, to be employed in limited quantum computers, this makes the development of suitable approximations particularly appealing. One way to design such approximations is to rely on classical computers to ``correct'' for correlations that low-rank UCC ans\"{a}tze might otherwise miss. To this end, recent works by the authors document the success of such an approach.\cite{windom2024new, windom2024attractive, windom2025toward} In the context of hybrid quantum-classical computing, it means relying on a quantum device to prepare a low-rank UCC state and carry out the subsequent optimization of the ansatz amplitudes and then relaying this information to a classical computer to compute corrections associated with the neglected higher-rank cluster operators.

Bearing in mind the limitations of current and near-term quantum computers, algorithms such as the variational quantum eigensolver (VQE)\cite{peruzzo2014variational} are better positioned to take advantage of current developments as they outsource part of the computational workload to classical computers. In this framework, the quantum computer is responsible for preparing a quantum state using a parameterized unitary operator--such as $e^\tau$ in UCC--, and estimating the expectation value of the system Hamiltonian. The classical computer contributes to this workflow by searching for an optimal set of parameters, i.e., the cluster amplitudes $\tau$, that minimize the sought-after expectation value. Arguably, the main difficult in this type of workflow is access to robust quantum computers, which is alleviated by assigning some of the computational tasks to the classical computer, facilitating scientific exploration that would otherwise be inaccessible to current near-term quantum computers if working based on exclusively quantum algorithms, e.g., quantum phase estimation.\cite{abrams1997simulation,abrams1999quantum}

Even when resorting to a hybrid algorithm, the current quantum hardware constraints still demand that states be prepared using efficient ans\"{a}tze, which has been a major motivation for the current and related works. Starting from either a UCC ans\"{a}tze that explicitly considers only double (UCCD) or single and double (UCCSD) excitations, we recently derived a series of corrections that consider the leading-order effects of missing singles and triples excitation operators, respectively.\cite{windom2024attractive,windom2024new} Our triples correction appeals to the fourth-order UCCSD energy functional, which we refer to as [T], and is particularly relevant to the current work.\cite{windom2024new} The resulting method, UCCSD[T], was shown to significantly improve upon baseline UCCSD results with respect to FCI. However, this method has not been assessed in relation to the corresponding ansatz that explicitly accounts for infinte-order triple excitation operators, namely UCCSDT. Furthermore, the behavior of additional terms in the sequence of perturbative triples corrections has not been studied in this context.
ork is two-fold. First, we derive a fifth-order triples correction to UCCSD which we call [T-5]. Our original intention was to provide a superior method to UCCSD[T], but as we discuss below (Section \ref{sec:results}), the meaning of ``superior'' is somewhat subjective. Second, we assess the performance of this newly developed approach in contrast with our previous fourth-order (UCCSD[T]) and infinite-order (UCCSDT) triples for the problem of N$_2$ and LiF dissociation. Our intention is to highlight any benefits that these triples corrections might yield as compared to their infinite-order, more demanding analog UCCSDT.
In situations where effects derived from triple excitations need to be accounted for, but explicit introduction of the underlying operators in the ansatz/circuit is cost prohibitive, these schemes would be a viable way to reduce the necessity of costly and scarce quantum computers while simultaneously improving the quality of low-rank ans\"{a}tze using classical resources.

\section{Theory}
\label{sec:theory}

Electron correlation energy, denoted by $\Delta E$, is generally defined as what is missing by the Hartree-Fock (HF) description of the wavefunction, i.e., $\Delta E = E-E_\text{HF}$, with the following corresponding Schr\"{o}dinger equation

\begin{equation}
    H_N\ket{\Psi}=\Delta E\ket{\Psi},
\end{equation}
and the normal-ordered Hamiltonian $H_N$
\begin{equation}
\begin{split}
\label{eq:ham}
H_N&=H-\braket{0|H|0}\\
&=\underbrace{\sum_{p}\epsilon_{pp}\{p^{\dagger}p\}}_{\text{$H_0$}}+ \underbrace{\frac{1}{4}\sum_{pqrs}\braket{pq||rs}\{p^{\dagger}q^{\dagger}sr\}}_{\text{$V$}}, \\
%& =  H_0 + V,
  \end{split}  
\end{equation}
where $p^\dagger$ and $p$ are creation and annihilation operators that act on the spin-orbital indexed by $p$. The real scalars $\epsilon_{pp}$ and $\braket{pq||rs}$ are one- and two-electron integrals, respectively. The mean-field reference function $\ket{0}$ is the canonical Hartree-Fock reference determinant hereafter.

For the overwhelming majority of examples in quantum chemistry, an exact solution to the time-independent Schr\"{o}dinger equation is algorithmically unfeasible, which makes the development of accurate and tractable ans\"{a}tze paramount. Fortunately, low-rank coupled cluster (CC) theory and its various flavors are known to recover a majority of correlation effects for most ``well-behaved'' systems, i.e., systems whose wavefunctions are dominated by a single mean-field configuration.\cite{bartlett1981many,bartlett1989coupled} The current work specifically focuses on unitary coupled cluster theory, which is characterized by the following exponential parameterization of the wavefunction 
\begin{equation}\label{eq:untrunatedUCCansatze}
    \ket{\Psi}_\text{UCC}=e^{\tau}\ket{0},
\end{equation}
where the exponential of the anti-Hermitian cluster operator, $\tau$, acts on the mean-field reference function $\ket{0}$. In the limit where $\tau$ accounts for  all possible de-excitations/excitations in an $N$-electron system
\begin{equation}
\begin{split}
        \tau &= \sum_n^N \tau_n \\
        \tau_n &= T_n - T_n^{\dagger} \\
        & = \frac{1}{(n!)^2}\sum_{ab\cdots ij\cdots} \big(t_{ij\cdots}^{ab\cdots}\{a^{\dagger}b^{\dagger}ij\}-t_{ab\cdots}^{ij\cdots}\{i^{\dagger}j^{\dagger}ab\}\big),
\end{split}
\end{equation}
the UCC ansatz is an exact parameterization of the wavefunction.\cite{evangelista2019exact} Note that $T_n$ and its adjoint are the regular cluster operators characterizing conventional CC theory. Since an exact parameterization is ultimately intractable, the maximal rank of the cluster operator must be truncated in practice. Although not typically  an issue when investigating ``well-behaved" molecular systems, there are several pathological counter-examples where low-rank CC/UCC is capable of achieving a desired threshold for accuracy.\cite{chan2004state} A traditional response to this issue has been to cheaply estimate the correlation effects of neglected cluster operators via perturbation theory, which is a well-documented way to increase agreement with FCI.\cite{urban1985towards,bartlett1990non,kucharski1998noniterative,kucharski1989coupled,kucharski1993coupled,raghavachari1989fifth}

To this end, the following subsections examine the UCC energy functional at fourth- and fifth-order in many-body perturbation theory (MBPT). Manipulations of the fourth-order energy functional recover the previously proposed [T] correction to UCCSD,\cite{windom2024new} and further supplements analysis of the fifth-order UCC energy functional used to define the [T-5] correction defined below.

\subsection{Fourth-order triples corrections to UCCSD}
\label{ssec:fourth_order}

For brevity, we will express the fourth-order UCC energy functional as
\begin{equation}\label{eq:UCC4orig}
    \Delta E^{[4]} = E^{[4]}(\text{SD}) + \Delta E^{[4]}(\text{T}),
\end{equation}
where $E^{[4]}(\text{SD})$ represents all fourth-order terms in UCCSD, the superscript denotes a generalized order in MBPT, and $\Delta E^{[4]}(\text{T})$ is defined as
\begin{equation}\label{eq:spelledOut4}
    \Delta E^{[4]}(\text{T}) = \braket{0|T_3^{\dagger}f_NT_3+T_2^{\dagger}W_NT_3+T_3^{\dagger}W_NT_2|0}.
\end{equation}
Varying Equation \ref{eq:spelledOut4} with respect to $T_3^{\dagger}$ yields the residual equations Equation \ref{eq:residual}
\begin{equation}
\label{eq:residual}
    \frac{\partial \Delta E^{[4]}(\text{T})}{\partial T^\dagger _3} = f_NT_3 + W_NT_2 = 0,
\end{equation}
which supports the creation of an approximate $T_3$ that is correct through second-order in MBPT
\begin{equation}\label{eq:SOt3}
\begin{split}
        T_3^{[2]} = R_3\bigg(W_NT_2\bigg).
\end{split}
\end{equation} We note that $R_n(X)$ is the resolvent operator 
\begin{equation}
\label{eq:resolvent}
\begin{split}
    &R_n(X) = \\
    &(n!)^{-2}\sum \frac{\braket{\Phi_{ij\cdots}^{ab\cdots}|X|0}}{\epsilon_i+\epsilon_j +\cdots - \epsilon_a - \epsilon_b-\cdots} \{a^{\dagger}b^{\dagger}\cdots j i \},
\end{split}
\end{equation}
which modulates projections onto particular excitation manifolds thereby ensuring pertinent denominators are used. Equation \ref{eq:SOt3} is further illustrated diagrammatically by a1 in Figure \ref{fig:Tcorr}.

Upon insertion of Equation \ref{eq:SOt3} into Equation \ref{eq:spelledOut4}, we recover the final contribution to the fourth-order energy associated with triple excitations:
\begin{subequations}
    \begin{align}
        &\Delta E(\text{T}) = \braket{0|T_2^{\dagger}W_NT_3|0},\\
        \begin{split}
        &\Delta E^{[4]}(\text{T}) = \\
        &\braket{0|T_2^{\dagger}W_NR_3(W_NT_2)|0} \equiv \braket{0|T_2^{\dagger}W_NT_3^{[2]}|0}. \label{eq:ucc4expression}
        \end{split}
    \end{align}
\end{subequations}
Equation \ref{eq:ucc4expression} formally defines the previously proposed [T] correction to account for the leading-order effects of triple excitations associated with the UCCSD ansatz.\cite{windom2024new} Diagram a2 of Figure \ref{fig:Tcorr} illustrates the diagrammatic interpretation of [T]. 

\subsection{Fifth-order triples corrections to UCCSD}

Recognizing that the equations found at fourth-order (Section \ref{ssec:fourth_order}) are subsumed within those found at higher orders, the cumulative fifth-order energy functional  can be written such that:
\begin{equation}
    \label{eq:functional_5}
    \Delta E(5) = E^{[5]}(\text{SD}) + \Delta E^{[5]}(\text{T}) + \Delta E^{[4]},
\end{equation}
where $E^{[5]}(\text{SD})$ accounts for all terms exclusively involving singles and doubles, $\Delta E^{[5]}(\text{T})$ contains all fifth-order contributions to the energy involving $T_3$, $\Delta E^{[4]}$ is given by Equation \ref{eq:UCC4orig}, and the $\Delta E(5)$ represents the overall, cumulative contribution to the fifth-order functional that explicitly includes the [T] correction found at fourth-order as expressed in Equation \ref{eq:ucc4expression}.  Note that we are particularly interested in $\Delta E^{[5]}(\text{T})$, which is defined as 
\begin{equation}
    \label{eq:fifth_t_functional}
    \begin{split}
     \Delta E^{[5]}&(\text{T}) =\\ 
     &T_3^{\dagger}W_NT_3 + \frac{1}{2}\bigg( (T_2^{\dagger})^2W_NT_3 + T_3^{\dagger}W_NT_2^2\bigg).       
    \end{split}
\end{equation}
Varying the cumulative fifth-order energy functional defined in Equation \ref{eq:functional_5} with respect to $T_3^{\dagger}$ leads to a (cumulative) fifth-order contribution to the underlying residual equation
\begin{equation}\label{eq:residual_5}
        \frac{\partial \Delta E(\text{5})}{\partial T_3^{\dagger}} = f_NT_3+W_NT_2 + W_NT_3 + \frac{1}{2}W_NT_2^2 = 0. 
\end{equation}

\begin{figure}[ht!]
%\centering
\includegraphics[width=\columnwidth]{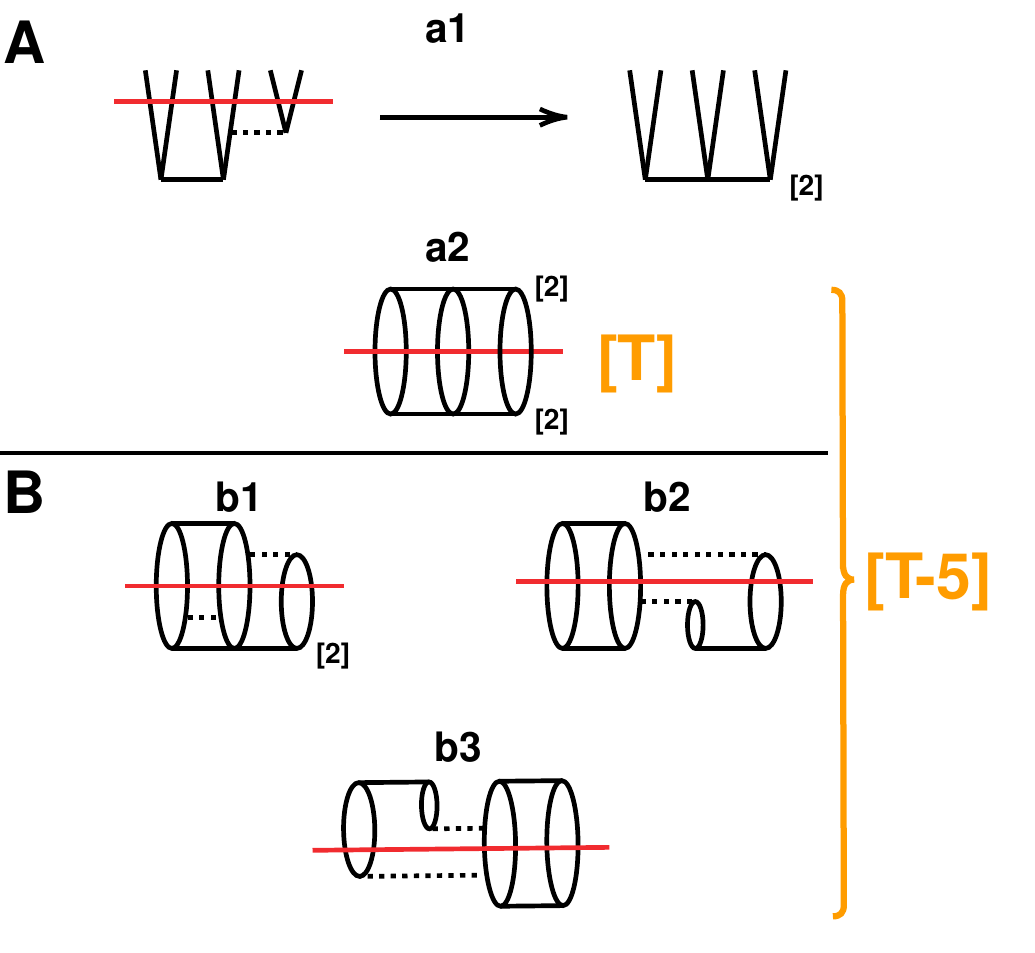}    

\caption{\label{fig:Tcorr}
Illustration of the skeleton diagrams defining the fourth- ([T]) and fifth-order ([T-5]) corrections to estimate the effect of neglected $\tau_3$ excitation operators in the UCCSD ansatz.  Red lines indicate a 6-index Fock denominator of the kind $\epsilon_{ijk}^{abc} = (\epsilon_i+\epsilon_j+\epsilon_k-\epsilon_a-\epsilon_b-\epsilon_c)^{-1}$, solid black lines indicate a cluster operator $T_n$ with $n$ vertices, the dotted lines indicate the two-body part of the Hamiltonian, and the subscript [2] indicates an amplitude that is correct through second-order in MBPT. The diagram a1 defines the approximated $T_3$ amplitude used in the [T],\cite{windom2024new} [T-5] (present work), and [Q-6]\cite{windom2025toward} methods, whereas diagram a2 represents the [T] energy correction. Incorporating diagrams b1-b3 on top of [T] leads to the [T-5] correction for triple excitations that is correct through fifth-order in MBPT.}

\end{figure}

Similarly with Equation \ref{eq:SOt3}, from Equation \ref{eq:residual_5} a pure third-order approximation to $T_3$ can be expressed as:
\begin{equation}\label{eq:TOt3}
    T_3^{[3]} = R_3\bigg( W_NT_3^{[2]} + \frac{1}{2}W_NT_2^2\bigg).
\end{equation}
Satisfaction of Equation \ref{eq:residual_5}, which leads to the second- and third-order approximations to $T_3$ found in Equations \ref{eq:SOt3} and \ref{eq:TOt3}, dramatically simplifies the fifth-order energy functional in Equation \ref{eq:fifth_t_functional}:
\begin{equation}
    \Delta E(5) = \braket{0|T_2^{\dagger}W_NT_3|0} + \frac{1}{2}\braket{0|(T_2^{\dagger})^2W_NT_3|0}.
\end{equation}
Recognizing that we are interested in the pure fifth-order contribution, we find that 
\begin{equation}
    \begin{split}
         \Delta E^{[5]}(\text{T}) &= \frac{1}{2}\braket{0|(T_2^{\dagger})^2W_NT_3^{[2]}+T_2^{\dagger}W_NT_3^{[3]}|0} \\
        & \equiv \frac{1}{2}\braket{0|(T_2^{\dagger})^2W_NR_3(W_NT_2)|0} \\
        & +\braket{0|T_2^{\dagger}W_NR_3(W_NT_3^{[2]} + \frac{1}{2}W_NT_2^2)|0}.
    \end{split}
\end{equation} These terms, in conjunction with the fourth-order correction defining [T], formally defines the cumulative fifth-order correction for neglected triples excitations which we refer to as [T-5]. Figure \ref{fig:Tcorr} illustrates this correction diagrammatically; diagram a2 in conjunction with b1-b3 define the [T-5] correction.

\section{Computational Details}
\label{sec:comp}

All UCC calculations are performed in the XACC software,\cite{mccaskey2020xacc, xacc_chem} using a numerical simulator which relies on PySCF\cite{sun2018pyscf} for one- and two-body molecular integrals that are employed in the construction of $\epsilon_{pp}$ and $\braket{pq||rs}$ in Equation \ref{eq:ham}. The STO-6G basis set is used throughout this work,\cite{Hehre1969,Feller1996,Pritchard2019} and we drop the core electrons in the correlation calculation. Once the $\tau$ amplitudes are obtained, the subsequent perturbative corrections are generated using the pyCC software.\cite{UT2} 

We investigate two flavors of the UCC ans\"{a}tze, namely the full operator, expressed as
\begin{equation}
  \label{eq:uccsd}
  |\Psi_\text{UCCSD}\rangle = e^{\sum_{ia}\theta_{i}^{a}(a^\dagger i- \text{h.c.})+\sum_{ijab}\theta_{ij}^{ab}(a^\dagger b^\dagger ji- \text{h.c.}))} |0\rangle
\end{equation}
and the trotterized variant, defined as
\begin{equation}
   |\Psi_\text{tUCCSD}\rangle = \prod_{ia}e^{\theta_i^a(a^\dagger i- \text{h.c.})}\prod_{ijab}e^{\theta_{ij}^{ab}(a^{\dagger}b^{\dagger}ji-\text{h.c.})}|0\rangle
  \label{eq:trotter}
\end{equation}

We note that, while the two ans\"{a}tze in Equations \ref{eq:uccsd} and \ref{eq:trotter} share similarities and the latter is presented as an approximation to the former, recent literature suggests that the trotterized version is its own unique ansatz construct.\cite{evangelista2019exact, Freericks2022} In both cases, the VQE is used to optimize the set of cluster amplitudes according to the criteria: 
\begin{equation}
   \tau_1^*,\tau_2^* = \argminA_{ \tau_1,\tau_2}\braket{\Psi(\tau_1,\tau_2)| H | \Psi( \tau_1,\tau_2)},
\end{equation}
which are subsequently decomposed into the corresponding $T$ and $T^\dagger$ and inserted into the equations derived in Section \ref{sec:theory}.

\section{Results and Discussion}
\label{sec:results}
\subsection{N$_2$ and LiF potential energy surfaces}

\begin{figure}[ht!]
\centering
\includegraphics[width=\columnwidth]{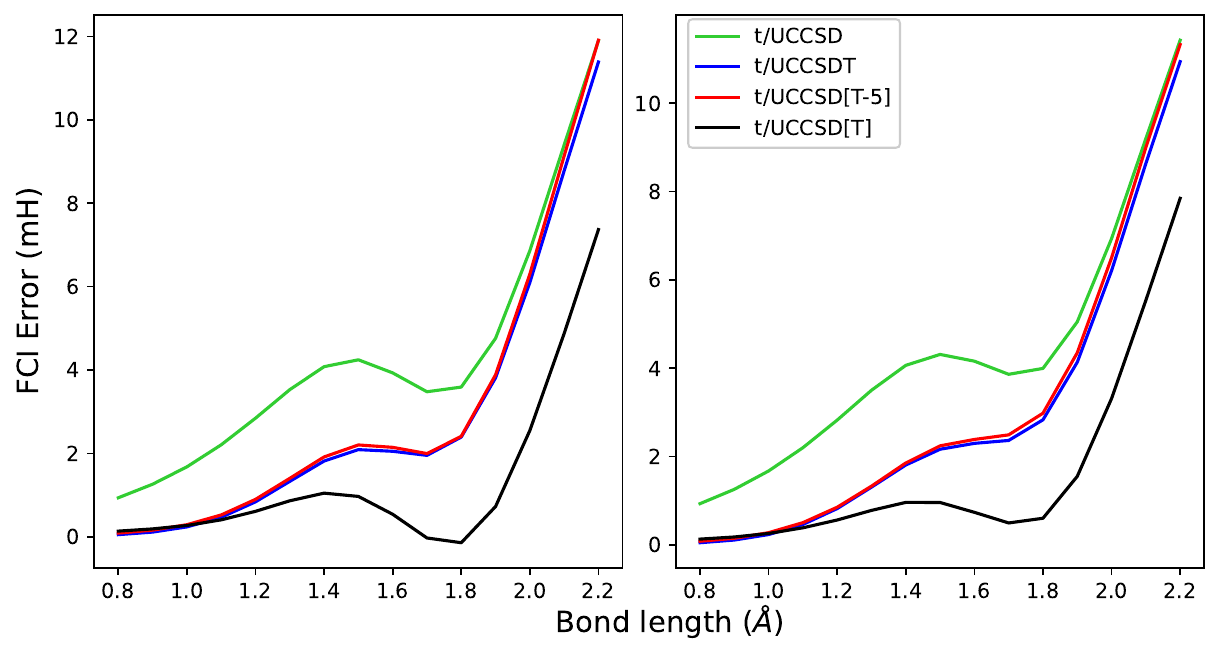}    

\caption{Errors with respect to FCI for the ground state energy of N$_2$ for UCCSD[T], UCCSD[T-5], and UCCSDT using the full (left) and trotterized (right) UCCSD ans\"{a}tze.}
\label{fig:FCIcmp}
\end{figure}

Figure \ref{fig:FCIcmp} compares the error in ground state energy estimation using the ansatz variants defined by Equations \ref{eq:uccsd} and \ref{eq:trotter} for UCCSD, UCCSD[T], UCCSD[T-5], and UCCSDT with respect to FCI for the dissociation of N$_2$. For bond lengths ranging from about 1.1 {\AA} to 2 {\AA}, UCCSD[T] is within chemical accuracy with respect to FCI. Perhaps most interesting is the fact that UCCSD[T] is superior to even UCCSDT over the reported range of the potential energy surface (PES). This result suggests that the [T] correction may benefit from fortuitous  cancellation of error that does not persist upon the inclusion of infinite-order triples into the ansatz. Albeit being somewhat counterintuitive, this is corroborated by the fact that a similar situation also arises in traditional CC theory where CCSD(T) tends to outperform CCSDT.\cite{stanton1997ccsd,bartlett2007coupled,watts1999equation}

\begin{figure}[ht!]
\centering
\includegraphics[width=\columnwidth]{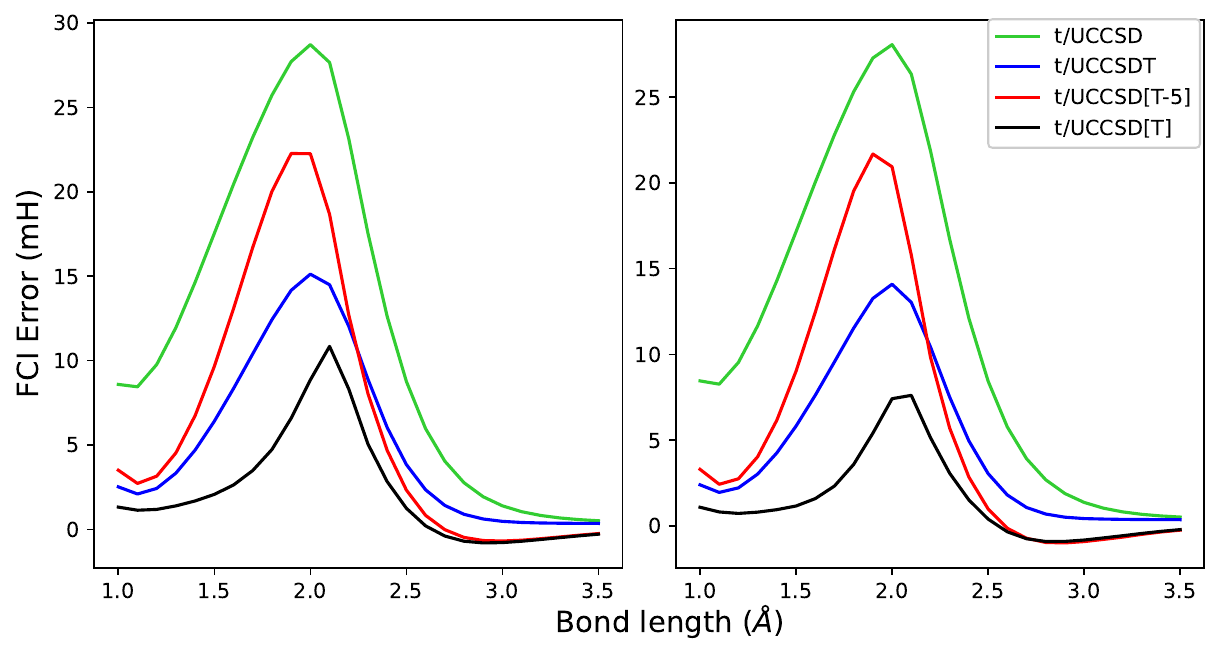}    

\caption{Errors with respect to FCI for the ground state energy of LiF for CCSD[T], UCCSD[T-5], and UCCSDT using the full (left) and trotterized (right) UCCSD ans\"{a}tze.}
\label{fig:LiFFCIcmp}
\end{figure}

These results are largely confirmed by Figure \ref{fig:LiFFCIcmp}, which compares the error of UCCSD, UCCSD[T], UCCSD[T-5], and UCCSDT in the dissociation of LiF. It is immediately apparent that including at least a partial accounting of triple excitation effects improves upon the performance of baseline UCCSD, but it manifests to varying extents for the different methods in question. Here again, we note that t/UCCSD[T] continues to offer the best agreement with respect to FCI. However, the set of fifth-order diagrams contributing to [T-5] leads to an overcorrection that worsens the agreement between t/UCCSD[T-5] and t/UCCSDT, particularly in the middle region of the PES.

\begin{figure}[ht!]
\centering
\includegraphics[width=\columnwidth]{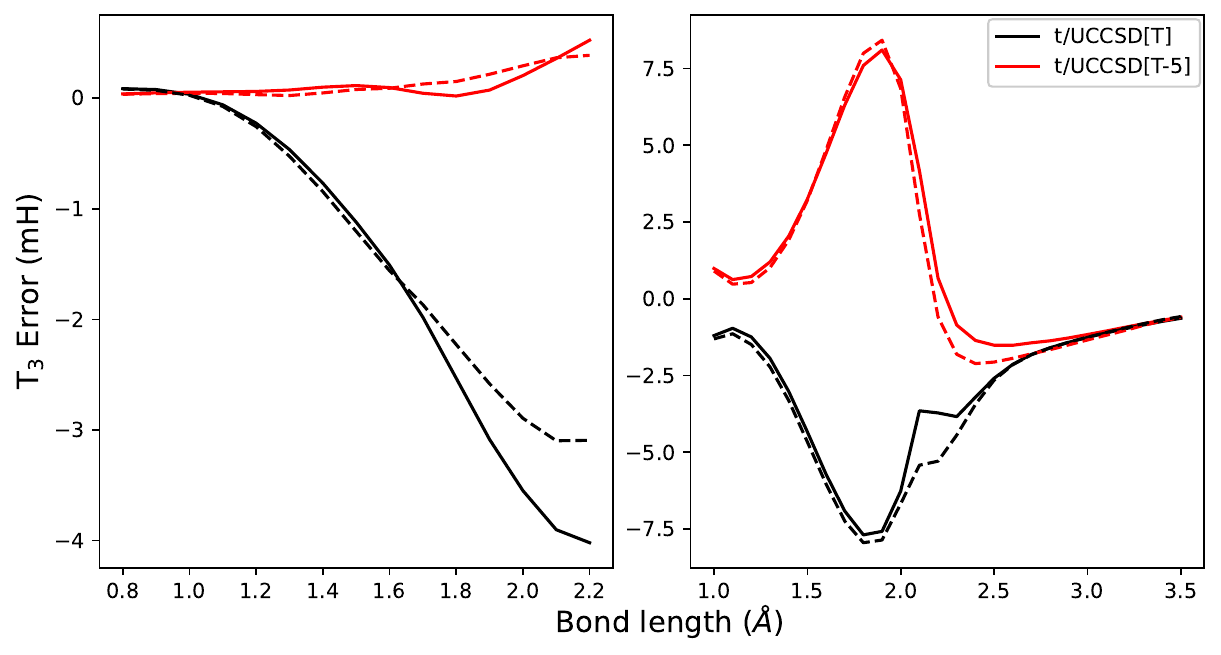}    

\caption{Comparison of fourth- ([T]) and fifth-order ([T-5]) triples corrections to full (solid) and trotterized (dashed) operator UCCSD with respect to the corresponding infinite-order UCCSDT across the N$_2$ (left) and LiF (right) PESs.}
\label{fig:T3cmp}
\end{figure}

Curiously, these general trends indicate that the [T-5] correction does not appreciably benefit from the addition of fifth-order contributions to triple excitations--at least in relation to FCI energetics. On the contrary, what we do find is that the resulting UCCSD[T-5]  method, at best, closely coincides with UCCSDT results. To examine this in more detail, Figure \ref{fig:T3cmp} juxtaposes the error of the UCCSD[T] and UCCSD[T-5] methods with respect to UCCSDT for the N$_2$ and LiF examples. We find that the [T-5] corrected methods are consistently within 1 mH of UCCSDT values for N$_2$ but overestimate this threshold by up to 8 mH for LiF. Moreover, the [T-5] consistently moves the energy estimate away from FCI and toward inclusion of full, infinite-order triples.

We note that one well-known downside to perturbative corrections in general is their degradation in stretched regions, which are characteristic behaviors of such corrections in situations where so-called ``strong" or static correlations dominate. Along with already being a recognized phenomenon in conventional CC theory, this artifact also has been seen and reported in our prior work on perturbative corrections to other flavors of UCC.\cite{windom2024new, windom2024attractive, windom2025toward} In fact, we have found that the trotterized ansatz (Equation \ref{eq:trotter}) is comparatively more sensitive to the so-called ``non-variational catastrophes'' than their full operator analogs. This appears to follow from the ambiguity in how to order the different cluster operators, which is exacerbated when away from the regime where a single configuration is dominant.\cite{grimsley2019trotterized,windom2024new} This effect also emerges in the current work for N$_2$, and as a consequence, we avoid showing portions of the trotterized operator PES beyond  2.2 {\AA} where the corresponding PT corrections are ill-behaved.

\subsection{Diagrammatic contributions} 

\begin{figure}[ht!]
\centering
\includegraphics[width=\columnwidth]{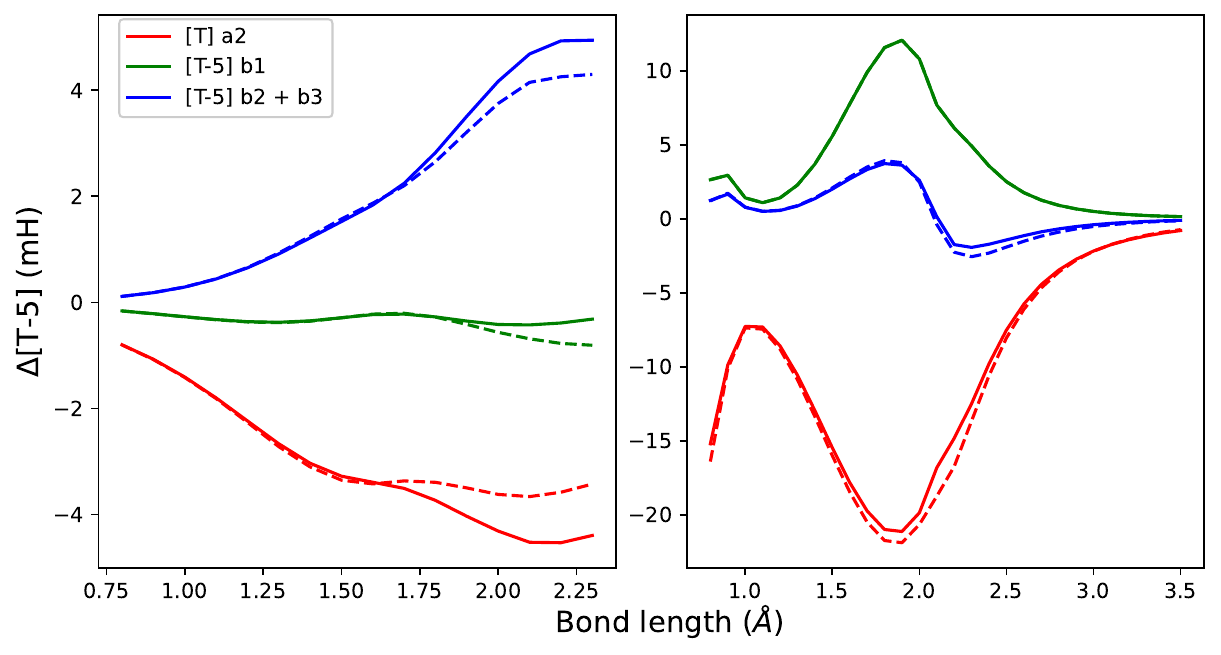}    

\caption{Individual contributions of diagrams in Figure \ref{fig:Tcorr} to the [T-5] correction for the full (solid) and trotterized (dashed) ans\"{a}tze across the N$_2$ (left) and LiF (right) PESs. We note that because diagrams b2 and b3 are Hermitian conjugates of one another, the corresponding line refers to their combined effect.}
\label{fig:DiagContrib}
\end{figure}

 %As there is an intimate relationship between MBPT and the exponential ans\"{a}tze, corrections to UCC should generally follow the behavior of corresponding MBPT terms.
 The linked diagram theorem establishes an intimate relationship between MBPT and the exponential ansatz that is characteristic of CC theory.\cite{shavitt2009many} In MBPT, pure even-order corrections should yield negative contributions to the correlation energy by virtue of the negative definite resolvent (Equation \ref{eq:residual}), whereas no such expectation is guaranteed for pure odd-order terms. As the corrections to UCC proposed in this work are informed by MBPT, the expectation is that the results should generally emulate MBPT, although strict equivalence is not guaranteed. To investigate this and to examine which term is responsible for the comparatively poor performance of the [T-5] correction, Figure \ref{fig:DiagContrib} decomposes its total contribution into individual, fourth- and fifth-order diagrammatic components as depicted in Figure \ref{fig:Tcorr} a2, b1, b2, and b3. 

For both N$_2$ and LiF, the [T] correction yields an overall negative correction--as illustrated in Figures  \ref{fig:FCIcmp} and \ref{fig:LiFFCIcmp}--,which is the tendency of even-ordered contributions as discussed above. In the case of LiF, this correction is actually quite large in general across the PES, increasing the agreement with FCI by up to 22 mH as compared to baseline t/UCCSD.  Although not as large in magnitude, the [T] nevertheless is also negative and increases agreement with respect to FCI for the N$_2$ PES by almost 5 mH.

On the other hand, the results for the pure fifth-order terms are less consistent. For N$_2$, the contribution of diagram b1 is consistently slightly below zero, whereas diagrams b2 + b3 are overwhelmingly positive and dominate at pure fifth-order. Conversely these trends are qualitatively reversed in the case of LiF, where diagram b1 now dominates at pure fifth-order and provides an overall positive contribution while diagrams b2 + b3 tend to be closer to zero on average. In either example, the comparatively large, positive fifth-order contribution tends to ``wash out'' some of the gains attributed to the pure fourth-order [T] correction. Although these insights are undoubtedly impacted by the finiteness of the basis set effects, these trends tend to indicate the importance of truncating perturbative corrections at even orders.  

\section{Conclusion}

This work derives fifth-order corrections that consider the effect of neglected triple excitation operators in the UCCSD ansatz leading to a method which we call UCCSD[T-5]. We then assess the performance of this method in tandem with the previously proposed UCCSD[T] and against infinite-order UCCSDT and FCI in a scan of the N$_2$ and LiF PESs. We find the [T] correction to be closest to FCI, and that the higher-order [T-5] correction actually makes the resulting UCCSD[T-5] behave similarly to UCCSDT--or worse. We call attention to the fact that, if the proposed approach were to be implemented with the aid of a quantum computer, both the [T] and [T-5] corrections to UCCSDperformed on a classical computer using amplitudes that would have been determined using the quantum device. In other words: the garnered gain in accuracy from these perturbative corrections requires no extra quantum resources to compute these triples corrections beyond what is required to implement the baseline UCCSD. 

These preliminary results suggest two points that are worthy of consideration. The first point recognizes the potential in adopting the [T] to augment UCCSD, which requires significantly less quantum resources than UCCSDT while increasing the predictive fidelity. The second point to consider is that if UCCSDT quality results are actually the target--for whatever reason--then it seems plausible that the addition of even higher-order corrections beyond [T-5] could, in principle, be a viable way of achieving this goal. 

\section{Acknowledgements}

We acknowledge support by the “Embedding Quantum Computing into Many-body Frameworks for Strongly Correlated Molecular and Materials Systems” project, which is funded by the U.S. Department of Energy (DOE), Office of Science, Office of Basic Energy Sciences, the Division of Chemical Sciences, Geosciences, and Biosciences. This research used resources of the Compute and Data Environment for Science (CADES) at the Oak Ridge National Laboratory, which is supported by the Office of Science of the U.S. Department of Energy under Contract No. DE-AC05-00OR22725. Z.W.W. acknowledges several helpful discussions with Dr. Rodney J. Bartlett of the Quantum Theory Project, University of Florida.  

\bibliography{main}

\end{document}